\documentclass{article}
\usepackage{spconf,amsmath,graphicx,hyperref}
\usepackage{algorithm}
\usepackage{amsfonts}

\title{BridgeCode: A Dual Speech Representation Paradigm for Autoregressive Zero-Shot Text-to-Speech Synthesis}
\name{
Jingyuan Xing\textsuperscript{1}$^{*}$, 
Mingru Yang\textsuperscript{1}$^{*}$, 
Zhipeng Li\textsuperscript{1}, 
Xiaofen Xing\textsuperscript{1}$^{\dagger}$, 
Xiangmin Xu\textsuperscript{2}
\thanks{$^{*}$Jingyuan Xing and Mingru Yang are co-first authors.}
\thanks{ $^{\dagger}$Xiaofen Xing is the corresponding author.}}

\address{ \textsuperscript{1} South China University of Technology, Guangzhou, China \\
\textsuperscript{2} Foshan University, Foshan, China
}

\begin{document}
\ninept
\maketitle
\begin{abstract}

Autoregressive (AR) frameworks have recently achieved remarkable progress in zero-shot text-to-speech (TTS) by leveraging discrete speech tokens and large language model techniques. Despite their success, existing AR-based zero-shot TTS systems face two critical limitations: (i) an inherent speed–quality trade-off, as sequential token generation either reduces frame rates at the cost of expressiveness or enriches tokens at the cost of efficiency, and (ii) a text-oriented supervision mismatch, as cross-entropy loss penalizes token errors uniformly without considering the fine-grained acoustic similarity among adjacent tokens. To address these challenges, we propose \textbf{BridgeTTS}, a novel AR-TTS framework built upon the dual speech representation paradigm \textbf{BridgeCode}. BridgeTTS reduces AR iterations by predicting sparse tokens while reconstructing rich continuous features for high-quality synthesis. Joint optimization of token-level and feature-level objectives further enhances naturalness and intelligibility. Experiments demonstrate that BridgeTTS achieves competitive quality and speaker similarity while significantly accelerating synthesis. Speech demos are available at https://test1562.github.io/demo/.
\end{abstract}
\section{Introduction}
\label{sec:intro}

Inspired by the remarkable success of large language models (LLMs) \cite{yang2025qwen3, yang2024qwen2, floridi2020gpt, touvron2023llama}, autoregressive (AR) frameworks have advanced diverse fields. Recent advances in zero-shot text-to-speech (TTS) \cite{wang2025spark, du2024cosyvoice2, du2024cosyvoice, casanova2024xtts, kim2023transduce, betker2023better} have demonstrated the effectiveness of leveraging discrete speech tokens \cite{wang2025tadicodec, defossez2022high, du2024funcodec, yang2023hifi} and AR language models for high-quality synthesis. As a promising paradigm for zero-shot TTS, autoregressive approaches can achieve human-level parity in terms of naturalness and intelligibility under zero-shot scenarios, making it an attractive research area.

\begin{figure}[t]
  \centering
  \includegraphics[width=1\linewidth]{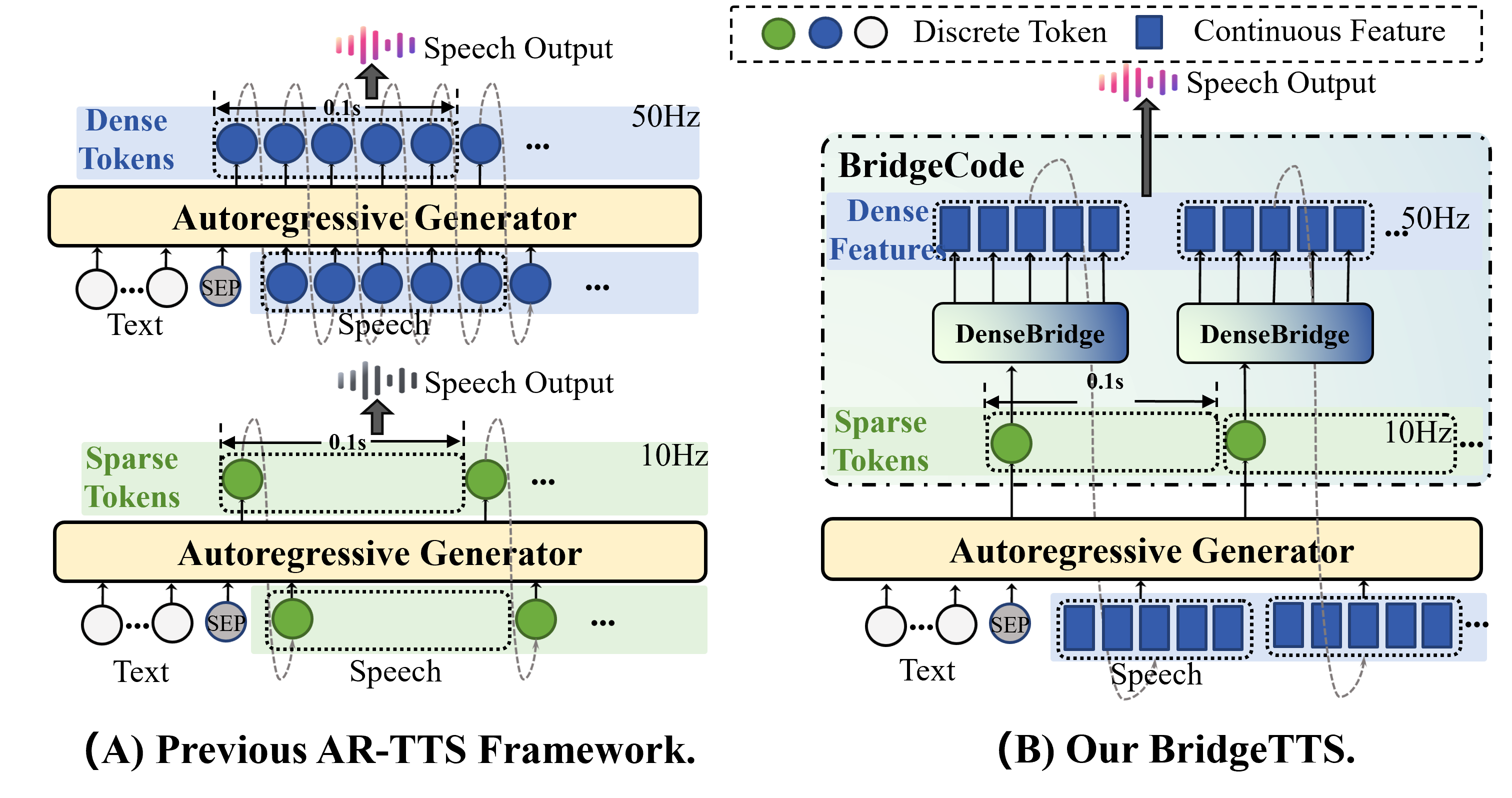}
  \vspace{-10pt}
  \caption{Comparison between existing AR-TTS frameworks and the proposed BridgeTTS. (A) Previous approaches exhibit an inherent trade-off between token generation rate and speech quality. (B) BridgeTTS employs sparse tokens for efficient AR generation and dense continuous features for high-quality synthesis via bridging module.}
  \label{fig: com}
\end{figure}

However, conventional AR zero-shot TTS systems still face two issues. First, the AR model generates discrete speech tokens sequentially, while a separate model decodes them into acoustic features and further synthesizes speech. This paradigm necessitates that the AR model iteratively processes token sequences with high generation rates, creating a computational bottleneck for real-time deployment. As illustrated in Fig.\ref{fig: com}(A), existing methods either reduce token rates (bottom of Fig.\ref{fig: com}(A)), sacrificing the expressiveness of generated speech \cite{wang2025tadicodec, ji2024wavtokenizer}, or enrich tokens with additional information (top of Fig.\ref{fig: com}(A)), reducing the token generation efficiency \cite{2024arXiv240605370C}. This leads to \textbf{issue (i): an inherent token rate-quality trade-off} between the token generation rate of AR and the quality of the synthesized speech. Second, existing AR zero-shot TTS methods \cite{du2024cosyvoice2} typically adopt training paradigms directly inherited from large language models, employing cross-entropy loss that solely focuses on token prediction accuracy. While this approach is well-suited for text token prediction, it is suboptimal for speech token generation. Adjacent speech tokens in the acoustic space often differ only in subtle prosodic or timbral variations, yet cross-entropy loss applies a uniform penalty regardless of token proximity to ground truth. This fails to provide the fine-grained, hierarchical supervision necessary for high-quality speech synthesis, leading to \textbf{issue (ii): text-oriented supervision mismatch}.

To address issue (i), we propose \textbf{BridgeTTS}, a novel AR-TTS framework that reduces AR iterations while maintaining synthesis quality under zero-shot scenarios. BridgeTTS incorporates a core component, \textbf{BridgeCode}, which encompasses dual speech representations: sparse tokens and dense continuous features, along with two bridging modules for bidirectional conversion between them. As illustrated in Fig. \ref{fig: com}(B), BridgeTTS enables the AR model to generate sparse tokens for efficiency, while the bridging module reconstructs detailed continuous features to ensure high-quality speech synthesis, thereby reducing prediction steps without compromising synthesis quality. Furthermore, to address issue (ii), BridgeTTS refines the training paradigm by jointly optimizing token-level and feature-level objectives, providing fine-grained, hierarchical supervision for speech token prediction, which is essential for generating speech with enhanced naturalness and intelligibility.

In summary, our contributions are as follows:

1) We propose BridgeCode, a dual speech representation paradigm that incorporates sparse tokens and dense continuous features, along with two trained bridging modules for bidirectional conversion between them.

2) Building on BridgeCode, we further introduce the BridgeTTS framework, which substantially reduces AR prediction steps without compromising synthesis quality.

3) Experimental results demonstrate that BridgeTTS attains the lowest AR token rate among existing methods while achieving competitive naturalness and speaker similarity, and effectively mitigating error accumulation while accelerating synthesis speed.




\begin{figure}[t]
  \centering
  \includegraphics[width=\linewidth]{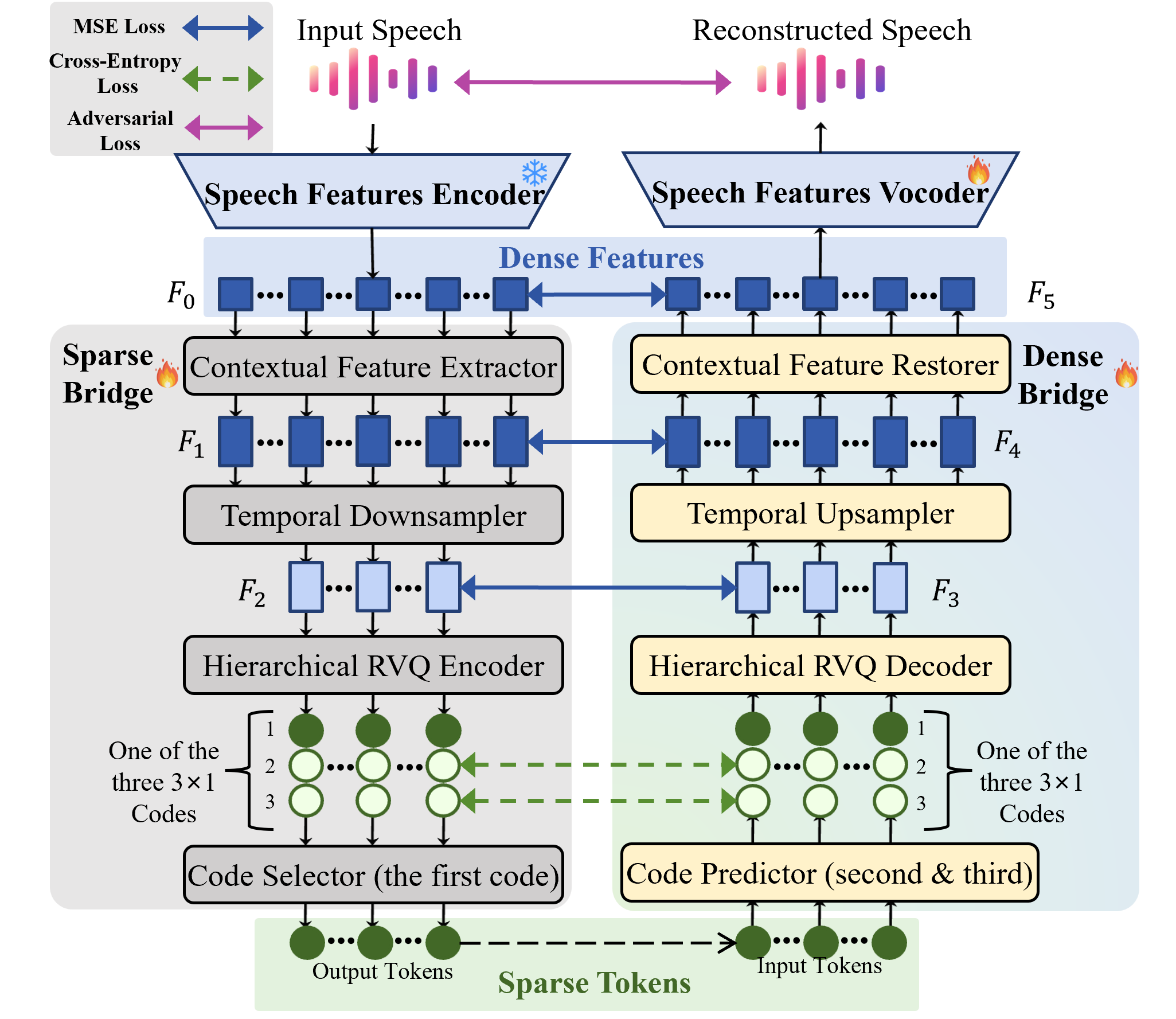}
  \vspace{-10pt}
  \caption{Overview of BridgeCode and architecture of bridging modules. Bidirectional arrows indicate loss constraints during training.}
  \label{fig:llm}
\end{figure}

\section{Method}
In this section, we describe our method in detail, including BridgeCode, a dual speech representation paradigm, and BridgeTTS, a novel AR-TTS framework built upon BridgeCode. 



\subsection{BridgeCode}


To establish the dual speech representations required by BridgeTTS, we propose BridgeCode, a dual speech representation paradigm that incorporates sparse tokens and dense continuous features with bridging modules between them. As illustrated in Fig.\ref{fig:llm}, dense continuous features are extracted by the frozen feature encoder from GPT-Talker \cite{liu2024generative}, while sparse tokens are obtained by compressing dense continuous features through the proposed SparseBridge. SparseBridge and DenseBridge are two symmetrical bridging networks that perform bidirectional conversion between these two representations. Moreover, to achieve fine-grained sparse-to-dense alignment, we draw inspiration from VDVAE \cite{child2020very} and enforce layer-wise alignment between intermediate features in both bridging modules (Fig.\ref{fig:llm}), ensuring high-fidelity bidirectional conversion between sparse tokens and dense continuous features. The network architecture and training objectives of SparseBridge and DenseBridge are detailed below. 


\textbf{SparseBridge Architecture.} SparseBridge compresses dense continuous features into sparse tokens while preserving essential information. Taking continuous speech features $F_0 \in \mathbb{R}^{T \times 768}$ as input, the contextual feature extractor employs multi-scale convolutional layers with kernel sizes of $1$, $3$, and $5$ to capture contextual dependencies across varying temporal spans. The extracted multi-scale features are then concatenated and denoted as $F_1 \in \mathbb{R}^{T \times 2304}$. To obtain sparse tokens, $F_1$ undergoes temporal downsampling to reduce the frame rate by a factor of $5$, yielding $F_2 \in \mathbb{R}^{T/5 \times 2304}$, which is then processed by hierarchical residual vector quantization (RVQ) \cite{barnes1996advances} that iteratively compresses $F_2$ into discrete codes. The Hierarchical RVQ Encoder processes the $2304$-dimensional vector by splitting it into three $768$-dimensional vectors, each quantized by a $3$-level RVQ, resulting in a $3 \times 3$ code matrix. Since VALL-E \cite{wang2023neural} has demonstrated that only the first RVQ indices are crucial while other indices can be discarded without significant information loss, a code selector retains solely the first RVQ indices, producing a sparse token sequence that preserves essential speech information while achieving substantial compression.

\textbf{DenseBridge Architecture.} DenseBridge employs a symmetric architecture to SparseBridge, designed to reconstruct dense continuous features from sparse tokens. Initially, a code predictor predicts the missing RVQ codes conditioned on sparse tokens, yielding complete RVQ indices. Subsequently, a hierarchical RVQ decoder progressively recovers quantized features from the complete code sequence. These features are then upsampled to restore the original temporal resolution, followed by a multi-scale convolutional inverse network that refines the features to reconstruct the continuous speech representation.

\textbf{Training and optimization.} To ensure precise alignment during the compression-reconstruction process, we enforce layer-wise alignment between intermediate features in both bridging modules during training. The code prediction loss $\mathcal{L}_{\text{code}}$ and the feature reconstruction loss $\mathcal{L}_{\text{feat}}$ are employed to constrain the alignment between discrete tokens and continuous features at corresponding layers in DenseBridge and SparseBridge. As illustrated in Fig. \ref{fig:llm}, $\mathcal{L}_{\text{code}}$ is computed as the cross-entropy loss between the predicted second and third RVQ codes and the original codes. Meanwhile, $\mathcal{L}_{\text{feat}}$ is formulated as the mean squared error (MSE) loss between the reconstructed and compressed continuous features. Additionally, to ensure that dense features reconstructed by DenseBridge can generate high-quality speech, HiFi-GAN \cite{kong2020hifi} is employed as the vocoder to synthesize speech from the reconstructed dense features. The adversarial loss $\mathcal{L}_{\text{adv}}$ is computed using HiFi-GAN's  multi-scale and multi-period discriminators to minimize the difference between the reconstructed and original speech.
Therefore, the overall loss \( \mathcal{L}_{\text{total}} \) is formulated as follows:

\begin{align}
\mathcal{L}_{\text{total}} &= \mathcal{L}_{\text{code}} + \mathcal{L}_{\text{feat}} +\mathcal{L}_{\text{adv}}
\end{align}

\begin{figure*}[t]
  \centering
  \includegraphics[width=0.94\linewidth]{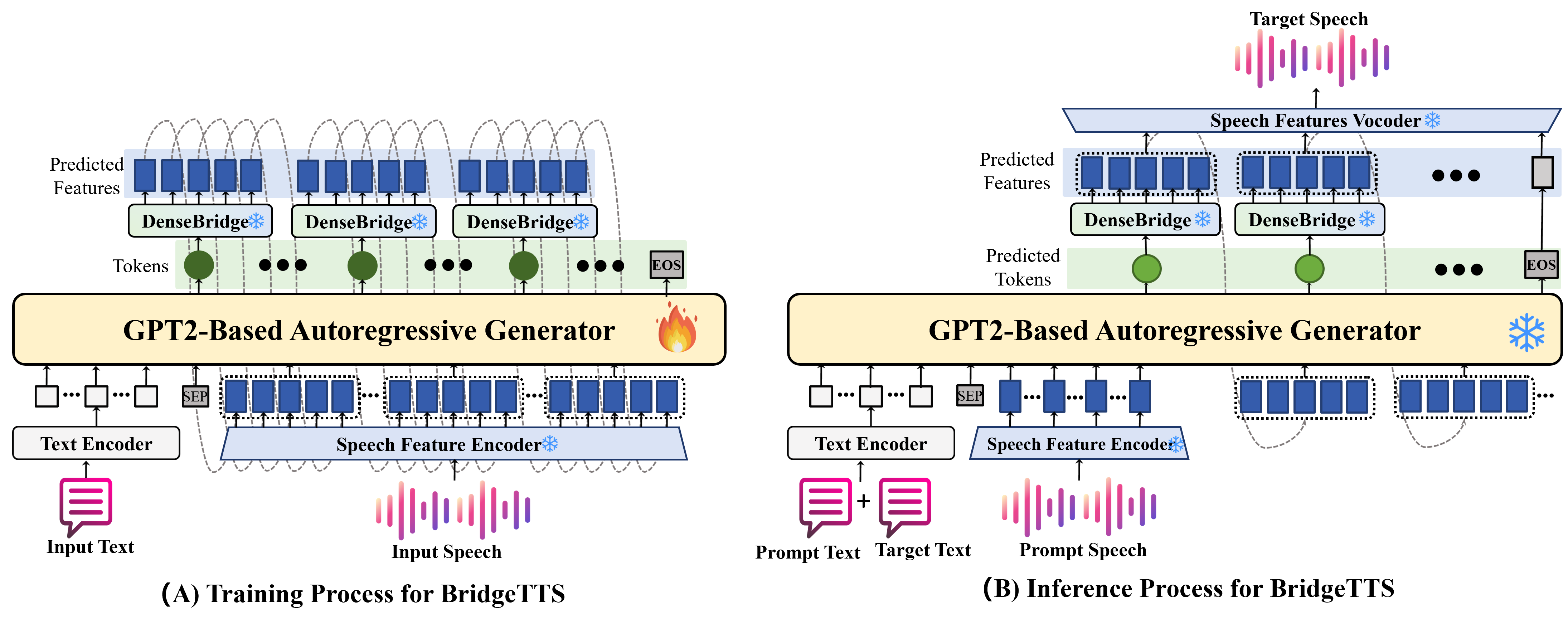}
  \vspace{-6pt}
  \caption{Overview of the proposed BridgeTTS. (A) Training Process Diagram. (B) Inference Process Diagram.}
  \label{fig:xunl}
\end{figure*}

\subsection{BridgeTTS}

Based on the proposed BridgeCode, dual speech representations can be obtained for any given speech, with bidirectional conversion facilitated by the trained bridging modules. In this section, we further present BridgeTTS. As illustrated in Fig.\ref{fig:xunl}, the autoregressive generator in BridgeTTS is a retrained GPT-2-based \cite{radford2019language} model. BridgeTTS allows the AR model to generate sparse tokens at a reduced frame rate to minimize iteration steps, while information-rich dense continuous features are reconstructed by the frozen DenseBridge for high-quality speech synthesis. This approach effectively addresses the aforementioned inherent speed-quality trade-off issue in AR-TTS. Moreover, to address the text-oriented supervision mismatch issue, BridgeTTS refines the training paradigm by jointly optimizing token-level and feature-level objectives, providing fine-grained supervision essential for high-quality speech synthesis.

Notably, unlike traditional autoregressive models that sequentially input one token to predict the next discrete token, BridgeTTS allows the AR model to input five consecutive speech feature frames at each step to predict the next token. This enables the AR model to make more informed predictions based on richer contextual information. Meanwhile, the AR model can observe the continuous features directly used for speech generation when predicting the next token, allowing it to adjust output tokens to control the synthesis of subsequent continuous features. Consequently, BridgeTTS enables the AR model to achieve better control over the naturalness and intelligibility of synthesized speech during both training and inference. 
The respective processes for training and inference are depicted in Fig.\ref{fig:xunl} (A) and (B). 

During the training process, both text input and corresponding speech features are used. We employ token loss to enforce the AR model's predicted sparse tokens to match the ground truth tokens extracted by SparseBridge at each prediction step. The token loss $\mathcal{L}_{\text{token}}$ is formulated as:
\[
\mathcal{L}_{\text{token}} = - \sum_{t=1}^{T} \log P({e}_t | \mathbf{f}_{<t}, \mathbf{t}_{\text{ref}}, \mathbf{t}_{\text{target}})
\]
where $P$ denotes the model's predicted probability distribution, $e_t$ is the ground truth token at time step $t$, $\mathbf{f}_{<t}$ represents all the previous features up to time step $t-1$, and $\mathbf{t}_{\text{ref}}$ and $\mathbf{t}_{\text{target}}$ are the reference and target text representations, respectively. In the BridgeTTS framework, each token prediction is conditioned on the previous features, reference text, and target text, while previous tokens do not directly influence the current prediction.

However, token loss computes prediction accuracy through cross-entropy loss, treating any mismatch between predicted and ground truth tokens as equally incorrect, regardless of their acoustic similarity. This poses a challenge in speech synthesis, as acoustically similar tokens may generate speech differing only in subtle acoustic details, such as prosodic variations and phonetic articulation. Consequently, a model predicting tokens closely resembling the ground truth receives the same penalty as one predicting acoustically distant tokens. To address this limitation, we introduce feature loss that computes the MSE between predicted and ground truth features, providing fine-grained, hierarchical supervision for the AR model. The feature loss is given by:
\[
    \mathcal{L}_{\text{features}} = \sum_{t=1}^{T} || \mathbf{f}_t - \hat{\mathbf{f}}_t ||_2^2
    \]
where $\mathbf{f}_t$ is the ground truth feature at time step $t$, and $\hat{\mathbf{f}}_t$ is the predicted feature obtained by passing the predicted token through the pre-trained DenseBridge.

The overall training loss $\mathcal{L}_{\text{AR}}$ for the AR generator is:

\[
\mathcal{L}_{\text{AR}} = \mathcal{L}_{\text{token}} + \mathcal{L}_{\text{features}}
\]

During inference, the AR model is conditioned on reference speech features and text input tokens, where the latter are derived by concatenating the reference transcript with the target text and encoding the result using a text encoder. At each autoregressive prediction step, the AR generator first predicts the next discrete token based on the accumulated speech features and text representations. The predicted token is then transformed into continuous speech features via the pre-trained DenseBridge. These continuous features are concatenated with the existing speech feature sequence and fed back into the AR generator for subsequent token prediction. This iterative process continues until an End-of-Speech (EOS) token is generated or the target sequence length is reached. Finally, the complete generated continuous speech features are synthesized into audio using the speech feature vocoder fine-tuned during BridgeCode training.

\section{Experiments}

\subsection{Dataset}

We conduct all experiments on the LibriTTS dataset \cite{zen2019libritts} with a sampling rate of 16 kHz. LibriTTS is a large-scale, multi-speaker English corpus comprising 585 hours of speech from over 2,300 speakers. For training, we combine the subsets train-clean-100, train-clean-360, and train-other-500. The development set is constructed by merging dev-clean and dev-other, while the test set consists of test-clean and test-other.

\subsection{Experimental Setup}

\textbf{Implementation Details.} For BridgeCode training, we first download the pre-trained weights for wav2vec 2.0 Base \cite{baevski2020wav2vec}\footnote{https://github.com/eastonYi/wav2vec}. The speech feature encoder is kept frozen, while two bridging modules are trained on the LibriTTS training set for \(700k\) steps on an NVIDIA A800 GPU with a batch size of 16. We employ the AdamW optimizer with an initial learning rate of \(1.0 \times 10^{-4}\), decayed by a factor of \(0.999^{1/8}\) per epoch. After training two bridging modules, they are frozen, and the autoregressive generator is subsequently trained for \(600k\) steps under the same conditions. 

\textbf{Subjective Evaluation Setup.} We conducted a subjective evaluation with 20 human raters, who were first trained on anonymized speech samples to familiarize them with the criteria for assessing speaker similarity and speech quality. Each rater scored randomly selected samples on a 5-point scale for both naturalness and similarity. Specifically, we performed a Speaker Similarity Mean Opinion Score (SMOS) test to measure speaker resemblance in zero-shot TTS synthesis and a Quality Mean Opinion Score (QMOS) test to evaluate overall naturalness. 

\textbf{Objective Evaluation Setup.} For objective evaluation, we measure UTMOS \cite{saeki2022utmos} to assess overall naturalness and quality, and Word Error Rate (WER) using a Wav2Vec 2.0-large-based ASR model \cite{baevski2020wav2vec} to evaluate word-level synthesis accuracy. To further assess model performance, we introduce an additional metric, Token rate, which measures the frequency at which each AR model outputs discrete speech tokens (i.e., tokens per second).

\subsection{Comparison with Existing Methods}

We compare our BridgeTTS with state-of-the-art (SOTA) methods on the LibriSpeech dataset. We choose four SOTA methods as our baseline, including VALL-E \cite{wang2023neural}, UnionAudio \cite{yang2023uniaudio}, a modified version of GPT-Talker \cite{liu2024generative} adapted for zero-shot synthesis, and CosyVoice \cite{du2024cosyvoice}. For fair comparison, all models are further trained on LibriSpeech using their original pre-trained weights as initialization. The results are presented in Table \ref{tab:tts_evaluation}, where GT denotes ground-truth speech.

\begin{table}[t]
\centering
\renewcommand{\arraystretch}{1.1}
\caption{Comparison Experiments on LibriTTS Development and Test Set. SMOS and QMOS scores are reported with 95\% confidence intervals. The best results are shown in \textbf{bold}, and the second-best are \underline{underlined}. }
\resizebox{1\linewidth}{!}{
\begin{tabular}{cccccc}
\hline
\textbf{Model} & \textbf{Token Rate $\left( \downarrow \right)$} & \textbf{WER $\left( \downarrow \right)$} & \textbf{SMOS $\left( \uparrow \right)$} & \textbf{QMOS $\left( \uparrow \right)$} & \textbf{UTMOS $\left( \uparrow \right)$} \\ \hline
\multicolumn{6}{c}{\textbf{LibriTTS Development Set}} \\ \hline
GT              & / & 2.3\% & 4.41 $\pm$ 0.11 & 4.41 $\pm$ 0.13 & 4.258 \\ \hline
UniAudio \cite{yang2023uniaudio} & 50Hz & 11.4\% & 3.81 $\pm$ 0.12 & 3.92 $\pm$ 0.09 & 3.676 \\ 
GPT-Talker \cite{liu2024generative} & 50Hz & \underline{5.9\%} & 3.78 $\pm$ 0.11 & 3.96 $\pm$ 0.12 & 3.693 \\ 
CosyVoice \cite{du2024cosyvoice} & 25Hz & 6.8\% & \textbf{4.13 $\pm$ 0.12} & \textbf{4.36 $\pm$ 0.12} & \textbf{4.253} \\ 
BridgeTTS (Ours)     & \textbf{10Hz} & \textbf{3.4\%} & \underline{4.07 $\pm$ 0.11} & \underline{4.15 $\pm$ 0.09} & \underline{4.050} \\ \hline
\multicolumn{6}{c}{\textbf{LibriTTS Test Set}} \\ \hline
GT   & / & 3.1\% & 4.33 $\pm$ 0.11 & 4.32 $\pm$ 0.09 & 4.275 \\ \hline
VALL-E \cite{wang2023neural}   & 50Hz & 18.5\% & 3.64 $\pm$ 0.12 & 3.49 $\pm$ 0.11 & 2.728 \\ 
UniAudio \cite{yang2023uniaudio} & 50Hz & 12.9\% & 3.62 $\pm$ 0.12 & 3.83 $\pm$ 0.15 & 3.663 \\ 
GPT-Talker \cite{liu2024generative} & 50Hz & 16.4\% & 3.78 $\pm$ 0.12 & 3.84 $\pm$ 0.09 & 3.566 \\ 
CosyVoice \cite{du2024cosyvoice} & 25Hz & \underline{8.0\%} & \textbf{4.12 $\pm$ 0.08} & \textbf{4.29 $\pm$ 0.11} & \textbf{4.148} \\ 
BridgeTTS (Ours)    & \textbf{10Hz} & \textbf{4.9\%} & \underline{4.01 $\pm$ 0.12} & \underline{4.11 $\pm$ 0.13} & \underline{3.894} \\ \hline \vspace{-10pt}
\end{tabular}
}
\label{tab:tts_evaluation}
\end{table}

As observed, BridgeTTS delivers competitive synthesis quality compared to four baseline methods across SMOS, QMOS, and UTMOS metrics, while achieving the lowest WER and Token Rate. Compared to GPT-Talker, which shares the same base model and training data, BridgeTTS improves speech naturalness and similarity while maintaining lower Token Rate and WER. This improvement stems from the proposed novel BridgeCode and unique AR paradigm, which enables the AR model to input multiple consecutive speech feature frames at each step for next-token prediction, facilitating more informed predictions based on richer contextual information. Additionally, DenseBridge losslessly converts sparse tokens from previous steps into dense continuous features for AR input in subsequent predictions, ensuring the stability of this AR paradigm.
Compared to CosyVoice, which employs updated codec models, advanced AR architectures, and larger training datasets, BridgeTTS achieves competitive speech naturalness and similarity while maintaining a lower WER. This superior performance is attributed to the lower frame rate of sparse tokens in the proposed BridgeCode. For synthesizing speech of equivalent duration, BridgeTTS requires fewer AR iterations than CosyVoice, resulting in reduced error accumulation during inference and higher word-level synthesis accuracy.

\begin{table}[t]
\centering
\renewcommand{\arraystretch}{1.1}
\caption{Ablation Study Results on LibriTTS Test Set.}
\resizebox{1\linewidth}{!}{
\begin{tabular}{l c c c c c }
\hline
\textbf{~~~~~~~~~~~~Model} & \textbf{Token Rate $\left( \downarrow \right)$} & \textbf{WER $\left( \downarrow \right)$} & \textbf{SMOS $\left( \uparrow \right)$} & \textbf{QMOS $\left( \uparrow \right)$}  & \textbf{UTMOS $\left( \uparrow \right)$}\\ \hline
BridgeTTS     & 10Hz & \textbf{4.9\%} & \textbf{4.01 $\pm$ 0.12 } & \textbf{ 4.11 $\pm$ 0.13} & \textbf{3.894} \\ \hline
-w/o DenseBridge   & 10Hz & 13.8\% & 3.74 $\pm$ 0.11& 3.74 $\pm$ 0.12& 3.443 \\ 
-w/o $\mathcal{L}_{\text{features}}$    & 10Hz & 7.1\% & 3.92 $\pm$ 0.13& 3.96 $\pm$ 0.12& 3.471 \\ \hline
\end{tabular}
}
\label{tab:eats_speech_ablation}
\end{table}

\begin{table}[t]
\centering
\renewcommand{\arraystretch}{1.1}
\caption{Comparison of Baseline AR Generator vs.  BridgeTTS on LibriTTS Test Set.}
\resizebox{1\linewidth}{!}{
\begin{tabular}{ccccccc}
\hline
\textbf{System} & \textbf{RTF $\left( \downarrow \right)$} & \textbf{Token Rate $\left( \downarrow \right)$} & \textbf{WER $\left( \downarrow \right)$} & \textbf{SMOS $\left( \uparrow \right)$} & \textbf{QMOS $\left( \uparrow \right)$} & \textbf{UTMOS $\left( \uparrow \right)$}\\ \hline
\textbf{Baseline AR} & 1× & 50Hz & 9.8\% & - & - & -\\ 
\textbf{BridgeTTS} & \textbf{0.37× }& \textbf{10Hz} &\textbf{ 4.9\%} & +0.12 & +0.09 & +0.43\\ \hline
\end{tabular}
}
\label{tab:comparison_systems}
\end{table}

\subsection{Ablation Study}



Ablation studies are conducted to assess the contributions of sequence compression and feature loss, as summarized in Table \ref{tab:eats_speech_ablation}. w/o BridgeCode denotes training the AR generator on compressed tokens alone, without employing BridgeCode, while w/o $\mathcal{L}_{\text{features}}$ denotes training with Bridge but excluding the feature loss.
The results show that compressing token sequences without BridgeCode degrades quality, as the AR generator lacks sufficiently informative prompts for accurate synthesis. Similarly, removing the feature loss results in suboptimal performance, since this objective enforces attention to fine-grained acoustic and temporal characteristics beyond semantic content.

To further demonstrate the acceleration effect of BridgeTTS, we compare a baseline AR generator with its DenseBridge-enhanced counterpart in Table \ref{tab:comparison_systems}. The results confirm that BridgeTTS achieves faster synthesis with improved real-time factor (RTF) while preserving comparable quality.

\section{Conclusion}
In this paper, we proposed BridgeTTS, which incorporates a novel BridgeCode and a unique AR paradigm. BridgeCode introduces dual speech representations together with two bridging modules, enabling the AR model to generate sparse tokens to minimize iteration steps, while reconstructing information-rich dense continuous features for high-quality speech synthesis. In addition, BridgeTTS refines the training paradigm of AR models by providing fine-grained supervision, mitigating the mismatch between text-based and speech-based large models.
Experiments and ablation studies verify the effectiveness of BridgeTTS. Moreover, BridgeCode can be generalized to arbitrary AR-TTS models, demonstrating strong generalization and promising potential for future applications.
\section{Acknowledgement}
This work was partly supported by Guangdong Basic and Applied Basic Research Foundation  (2025A1515011203), Guangdong Provincial Key Laboratory of Human Digital Twin (2022B1212010004) and Nansha Key Project under Grant (2022ZD011).

\bibliographystyle{IEEEbib}
\bibliography{refs}

\end{document}